\documentclass[RNAAS,twocolumn]{aastex63}

\shorttitle{Sample article}
\shortauthors{Calder\'{o}n Bustillo et al.}
\graphicspath{{./}{figures/}}
\usepackage[figuresright]{rotating}

\begin{document}

\title{Mapping the Universe Expansion: Enabling percent-level measurements of the Hubble Constant with a single binary neutron-star merger detection.}

\correspondingauthor{Juan Calder\'{o}n Bustillo, Samson H.W. Leong}
\email{juan.calderon.bustillo@gmail.com, samson32081@gmail.com}

\author{Juan Calder\'{o}n Bustillo}
\affiliation{Instituto Galego de F\'{i}sica de Altas Enerx\'{i}as, Universidade de
Santiago de Compostela, 15782 Santiago de Compostela, Galicia, Spain}
\affiliation{Department of Physics, The Chinese University of Hong Kong, Shatin, N.T., Hong Kong}
\affiliation{Monash Centre for Astrophysics, School of Physics and Astronomy, Monash University, VIC 3800, Australia}
\affiliation{OzGrav: The ARC Centre of Excellence for Gravitational-Wave Discovery, Clayton, VIC 3800, Australia}

\author{Samson H.W. Leong}
\affiliation{Department of Physics, The Chinese University of Hong Kong, Shatin, N.T., Hong Kong}

\author{Tim Dietrich}
\affiliation{Institut f\"{u}r Physik und Astronomie, Universit\"{a}t Potsdam, Karl-Liebknecht-Str. 24/25, 14776 Potsdam, Germany}
\affiliation{Max-Planck-Institute for Gravitational Physics (Albert-Einstein-Institute), Am M{\"u}hlenberg 1, 14476 Potsdam-Golm, Germany}

\author{Paul D. Lasky}
\affiliation{Monash Centre for Astrophysics, School of Physics and Astronomy, Monash University, VIC 3800, Australia}
\affiliation{OzGrav: The ARC Centre of Excellence for Gravitational-Wave Discovery, Clayton, VIC 3800, Australia}

%% Note that the \and command from previous versions of AASTeX is now
%% depreciated in this version as it is no longer necessary. AASTeX 
%% automatically takes care of all commas and "and"s between authors names.

%% AASTeX 6.3 has the new \collaboration and \nocollaboration commands to
%% provide the collaboration status of a group of authors. These commands 
%% can be used either before or after the list of corresponding authors. The
%% argument for \collaboration is the collaboration identifier. Authors are
%% encouraged to surround collaboration identifiers with ()s. The 
%% \nocollaboration command takes no argument and exists to indicate that
%% the nearby authors are not part of surrounding collaborations.

%% Mark off the abstract in the ``abstract'' environment. 

\begin{abstract}
The joint observation of the gravitational-wave and electromagnetic signal from the binary neutron-star merger GW170817 allowed for a new independent measurement of the Hubble constant $H_0$, albeit with an uncertainty of about 15\% at 1$\sigma$. Observations of similar sources with a network of future detectors will allow for more precise measurements of $H_0$. These, however, are currently largely limited by the intrinsic degeneracy between the luminosity distance and the inclination of the source in the gravitational-wave signal. We show that the higher-order modes in gravitational waves can be used to break this degeneracy in astrophysical parameter estimation in both the inspiral and post-merger phases of a neutron star merger. We show that for systems at distances similar to GW170817, this method enables percent-level measurements of $H_0$ with a single detection. This would permit the study of time variations and spatial anisotropies of $H_0$ with unprecedented precision. 
We investigate how different network configurations affect measurements of $H_0$, and discuss the implications in terms of science drivers for the proposed 2.5- and third-generation gravitational-wave detectors. Finally, we show that the precision of $H_0$ measured with these future observatories will be solely limited by redshift measurements of electromagnetic counterparts. 
\end{abstract}

%% Keywords should appear after the \end{abstract} command. 
%% See the online documentation for the full list of available subject
%% keywords and the rules for their use.
\keywords{Hubble constant --- 
Binary neutron-star mergers --- gravitational waves }

%% From the front matter, we move on to the body of the paper.
%% Sections are demarcated by \section and \subsection, respectively.
%% Observe the use of the LaTeX \label
%% command after the \subsection to give a symbolic KEY to the
%% subsection for cross-referencing in a \ref command.
%% You can use LaTeX's \ref and \label commands to keep track of
%% cross-references to sections, equations, tables, and figures.
%% That way, if you change the order of any elements, LaTeX will
%% automatically renumber them.
%%
%% We recommend that authors also use the natbib \citep
%% and \citet commands to identify citations.  The citations are
%% tied to the reference list via symbolic KEYs. The KEY corresponds
%% to the KEY in the \bibitem in the reference list below. 

\section{Introduction}

The joint detection of gravitational-wave (GW) and electromagnetic (EM) radiation from the binary neutron star (BNS) merger GW170817 is a milestone for astrophysics~\citep{TheLIGOScientific:2017qsa,GBM:2017lvd} that has already driven major leaps forward in a number of research areas. 
Among the many profound science outcomes, GW170817 provided a new, distance-ladder independent measure of the expansion of the Universe~\citep{Abbott:2017xzu,Coughlin:2019vtv,Dietrich:2020lps}, parameterized by the Hubble constant.
The number of confirmed and putative BNS candidates in the third observing run~\citep{Abbott:2020uma} of Advanced LIGO~\citep{TheLIGOScientific:2014jea} and Advanced Virgo~\citep{TheVirgo:2014hva}, and the planned sensitivity increase of current- and future-generation GW detectors~\citep{ALIGOplus,Punturo:2010zz,Reitze:2019iox}, indicates that we can expect a significant increase in both the number of detected BNS mergers, as well as signal-to-noise ratios of detected events.
Improvements in the high-frequency regime ($\gtrsim1$ kHz) will also lead to the first detection of the post-merger phase of BNS mergers~\citep{Martynov2019,OzHF}, a stage when matter effects play a significant role and most extreme densities are probed.

Determining the Hubble constant from joint GW-EM observations of BNS mergers relies on measuring the luminosity distance to the source from the GW signal and the redshift of the host galaxy from the EM counterpart ~(\citep{Schutz1986}; although see Refs.~\citep{Messenger2012, Taylor2012} for other methods).
A key limitation of the former, however, is the degeneracy between the effects on the GW signal produced by luminosity distance and the inclination of the binary. Here, we study how this degeneracy can be broken in the context of future GW detectors in two ways: (i) via the inclusion of higher-order modes (HMs) in GW parameter estimation, and (ii) by accessing the ratio of the two GW polarisations, known as plus ``+'' and cross ``$\times$'', using multiple detectors. For face-on binaries, we show that the inclusion of HMs leads to major improvements of the distance and inclination estimates, independently of the detector network configuration. For edge-on binaries, we find that a three-detector network that can constrain the polarisation ratio and sky-location of the binary is key to correctly estimate the distance, regardless of the usage or omission of HMs. In both cases, the $H_0$ measurement will \textit{not} be limited by our ability to infer the luminosity distance via GWs, but by the accuracy of the redshift measurement. With redshift-measurement improvements,\textcolor{black}{} $2\%$ level measurements of $H_0$ could be possible with the observation of \textit{a single} BNS located at~$\sim40$ Mpc, consistent with the distance of GW170817. 
We show that for unequal mass systems, these improvements can be achieved with the signal emitted during the inspiral phase alone, independent of whether there is matter in the system or not; i.e., the method works for binary systems containing neutron stars and/or black holes. 
For equal-mass systems, we show that inclusion of matter effects in the post-merger phase is key to improve distance estimates.

We note that percent-level measurements of $H_0$ could be performed in a five year time-frame making use of five second-generation detectors \citep{Chen2018}, namely the two Advanced LIGO detectors \citep{TheLIGOScientific:2014jea}, Advanced Virgo \citep{TheVirgo:2014hva}, KAGRA \citep{KAGRA}, and the forthcoming LIGO India \citep{India}. However, this relies on the combination of many observations, therefore assuming that $H_0$ is the same in all directions and distances; i.e., the Universe is statistically isotropic and homogeneous on the scales of interest. Our results help to improve this strategy as we point out that percent-level measurements with a single observation are possible. Therefore, \textcolor{black}{our method paves a way toward the study of} anisotropies \citep{Collins1986} or time variations of $H_0$ \citep{Wu1996}, to obtain a significantly better and more detailed understanding of the evolution history of our Universe. \\

\begin{figure*}
\includegraphics[width=0.48\textwidth]{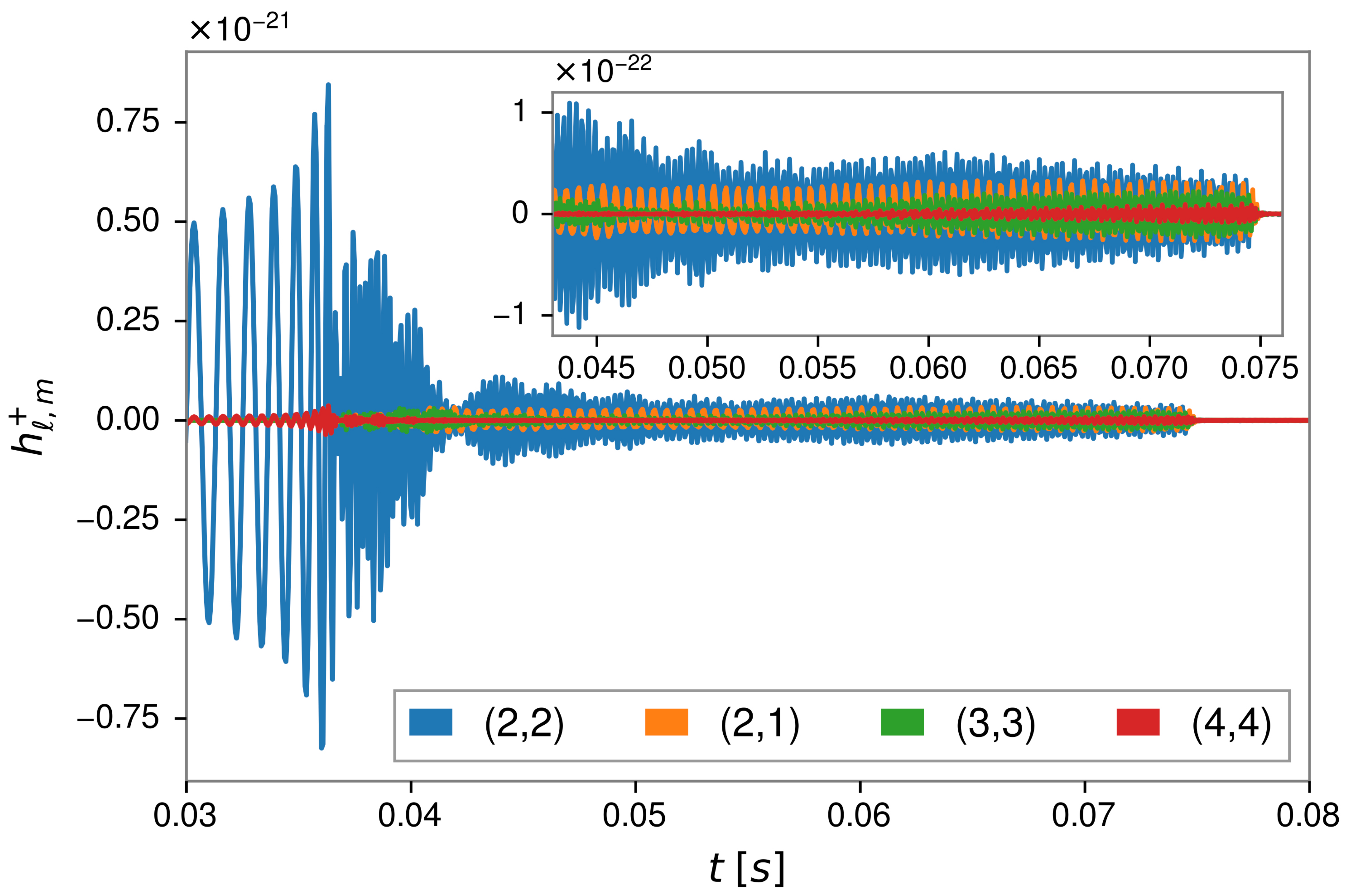}
\includegraphics[width=0.51\textwidth]{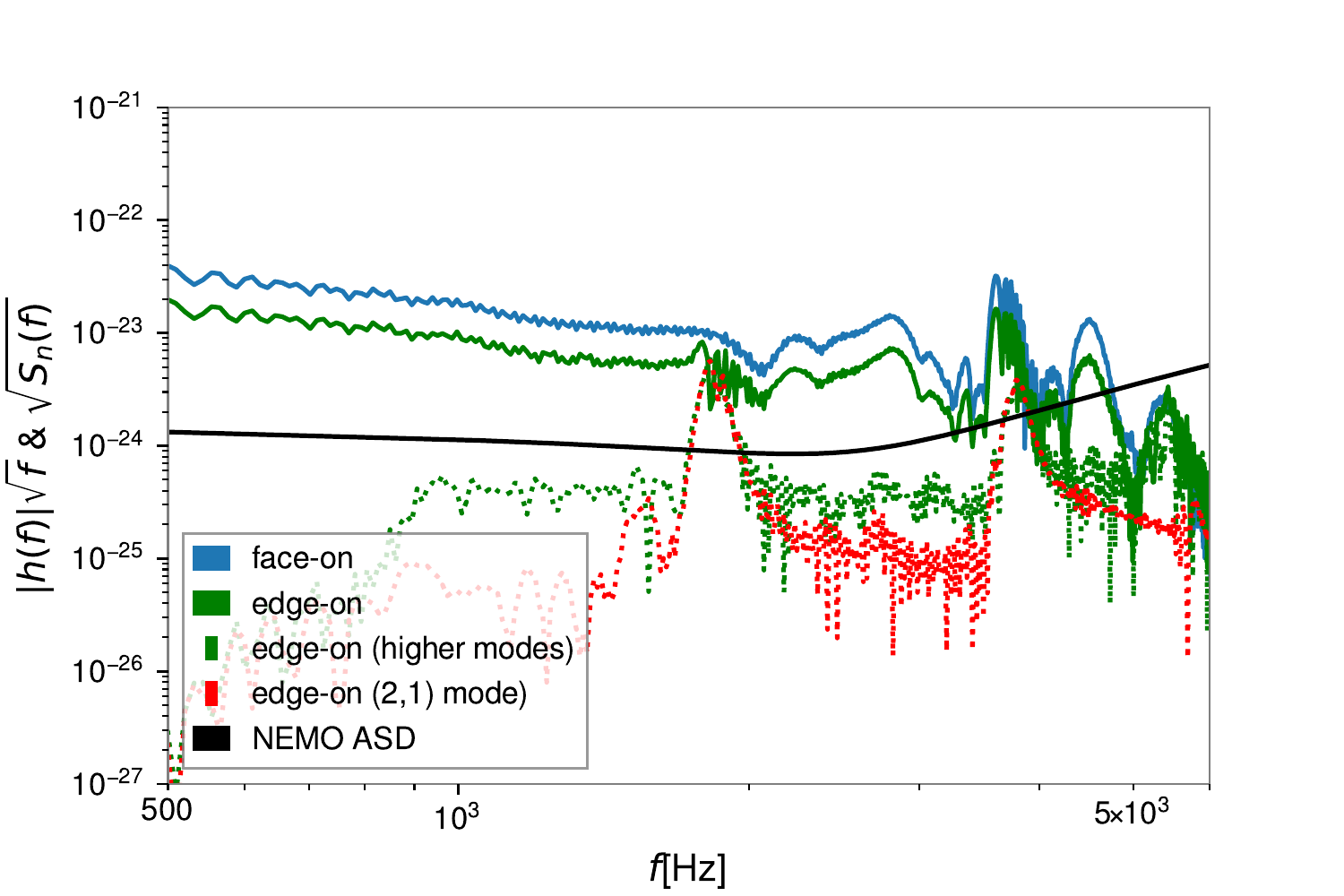}
\caption{GW modes and full spectrum of a binary neutron star merger and its post-merger remnant for the equal-mass \texttt{SLy} system used in this letter. The left panel shows the individual GW modes. The right panel shows the full spectrum of a signal observed face-on in blue, so that mainly the $(\ell,m)=(2,2)$ mode is present, and the signal observed edge-on. In addition, we show the contributions from the higher-modes and the $(\ell,m)=(2,1)$ mode. The black curve denotes the expected NEMO amplitude spectral density.} 
\label{fig:modes}
\end{figure*}

\subsection{\textbf{Higher order modes of compact binary mergers}}

\noindent The ``$+$'' and ``$\times$'' polarizations of a GW emitted by a compact binary merger located at a luminosity distance $d_L$ can be expressed as a superposition of individual modes, $h_{\ell,m}$, weighted by spin $-2$ spherical harmonics $Y^{-2}_{\ell,m}$ as 
\begin{equation}
h_+-ih_{\times} =\frac{1}{d_L}\sum_{\ell\geq 2}\sum_{m=-\ell}^{m=\ell}Y^{-2}_{\ell,m}(\iota,\varphi)h_{\ell,m}(\Xi).
\label{gwmodes}
\end{equation}
Here, $\Xi$ denotes the masses $m_i$ and dimensionless spins $\vec\chi_i$ of the individual objects and, for the case of BNSs, the individual tidal deformabilities $\Lambda_i$ characterizing the deformation of each star in the external gravitational field of the companion. The parameters $(\iota,\varphi)$ represent the polar and azimuthal angles of a spherical coordinate frame describing the location of the observer around the binary (or conversely, the orientation of the binary with respect to the observer), with $\iota=0$ denoting the direction of the orbital angular momentum of the binary and $\iota=\pi/2$ denoting the orbital plane. These values respectively refer to face-on and edge-on oriented binaries. For non-precessing binaries, the above sum is dominated by the quadrupole $(\ell,m)=(2,\pm 2)$ modes while higher-order modes become loud only during the final inspiral and merger phase, with increasing relative amplitude as the mass ratio $q=m_1/m_2 \geq 1$ increases~\citep{Pekowsky:2012sr, Varma:2014jxa,Bustillo:2015ova, Bustillo:2016gid}.\\

Current parameter estimation of BNS signals makes use of waveform templates including only the quadrupole mode. This causes a degeneracy between the inclination and distance parameters that fundamentally limits our ability to measure each. Several works have shown that inclusion of HMs in templates can break this degeneracy for sources with sufficiently-loud HMs in the detector sensitive band, as the observed combination of modes will depend on the orientation of the binary via the $Y_{\ell,m}$ factors~\citep{Graff:2015bba,London:2017bcn,CalderonBustillo:2018zuq,Pang:2018hjb,CalderonBustillo:2019wwe}. For Advanced LIGO and Virgo observations, unfortunately, this is only possible for large mass and asymmetric BBHs \citep{Graff:2015bba,London:2017bcn,Chatziioannou:2019dsz,LIGOScientific:2020stg} for which the merger and ringdown emission, rich in HMs, is strong in the detector sensitive band. In contrast, the merger and post-merger of BNSs is un-observable due to its large frequency. This emission will, however, be observable with future high-frequency~\citep{Martynov2019,OzHF} and third-generation~\citep{Punturo:2010zz,Reitze:2019iox} detectors and, in addition, the pre-merger GW emission will last for several minutes in the sensitive detector band, allowing to accumulate the effect of weak HMs.\\

\section{Analysis setup}

\subsection{\textbf{Binary neutron-star Waveforms}}

We test our ability to measure the source distance and $H_0$ using the inspiral and post-merger emission of BNSs. To this, we perform parameter inference on two kinds of simulated GW signals. First, we use 80 ms-long numerical-relativity simulations for the post-merger emission of  BNS~\citep{Dietrich:2017feu}. These have mass ratios $q=1$ and $q=1.5$ and implement two different equations of state (EOSs): a soft one (\texttt{SLy}) and a stiff one (\texttt{MS1b}~\footnote{While a stiff EOS like \texttt{MS1b} is disfavoured by the observation of GW170817 and its EM counterparts, e.g.,~\citep{GW170817Properties,Abbott:2018exr,Dietrich:2020lps}, it provides a good test case for our study to show the effect of two different EOSs.}). 
The left panel of Fig.~\ref{fig:modes} shows the time domain modes for the equal-mass \texttt{SLy} case located at a distance of $40$Mpc. The right panel shows in blue and green the spectra of full waveforms observed face-on (blue) and edge-on (green), together with the contribution of the HMs and the $(\ell,m)=(2,1)$ mode alone for the edge-on case. It can be noted not only how the face-on signal is stronger, but how the presence of HMs in the edge-on signal leads to noticeable morphological differences. Parameter inference on these short waveforms provides an idea of how the post-merger emission breaks the distance-inclination degeneracy. Restricting to this and ignoring the long inspiral signal, however, would greatly underestimate the signal-to-noise ratio (SNR) accumulated throughout the full minutes-long signal observable by the detector, reducing the accuracy of our measurements.\\

To obtain signals that can cover the full inspiral and post-merger emission from a BNS, we combine \citep{Bustillo:2015ova} our short numerical-relativity waveforms~\citep{Dietrich:2017feu} with long analytical waveforms covering the early-inspiral minutes before the merger (e.g., computed using tidal effective-one-body model of Ref.~\citep{Nagar:2018zoe}). Unfortunately, parameter inference using these waveforms is computationally prohibitive. As a solution, we implement a two-step approach. First, we consider 128s-long phenomenological (Phenom) waveforms~\citep{Khan:2015jqa,Santamaria:2010yb}, constructed with the \texttt{IMRPhenomHM} model~\citep{London:2017bcn}, covering the inspiral-merger and ringdown stages of non-precessing binary black hole (BBH) mergers. While computationally inexpensive, these waveforms omit the two main characteristic aspects of BNSs: tidal-deformability effects and the post-merger emission shown in Fig.~\ref{fig:modes}. Since, as argued, the latter can improve our distance measurements, results obtained using this BBH signal model are rather conservative. Finally, in order to obtain improved and more realistic results for BNSs, we combine these BBH parameter estimates with those obtained by solely analysing the short numerical-relativity waveforms covering the post-merger stage of BNSs, following the procedure described in~\citep{Zimmerman:2019wzo}.\\

\begin{figure*}
\includegraphics[width=0.49\textwidth]{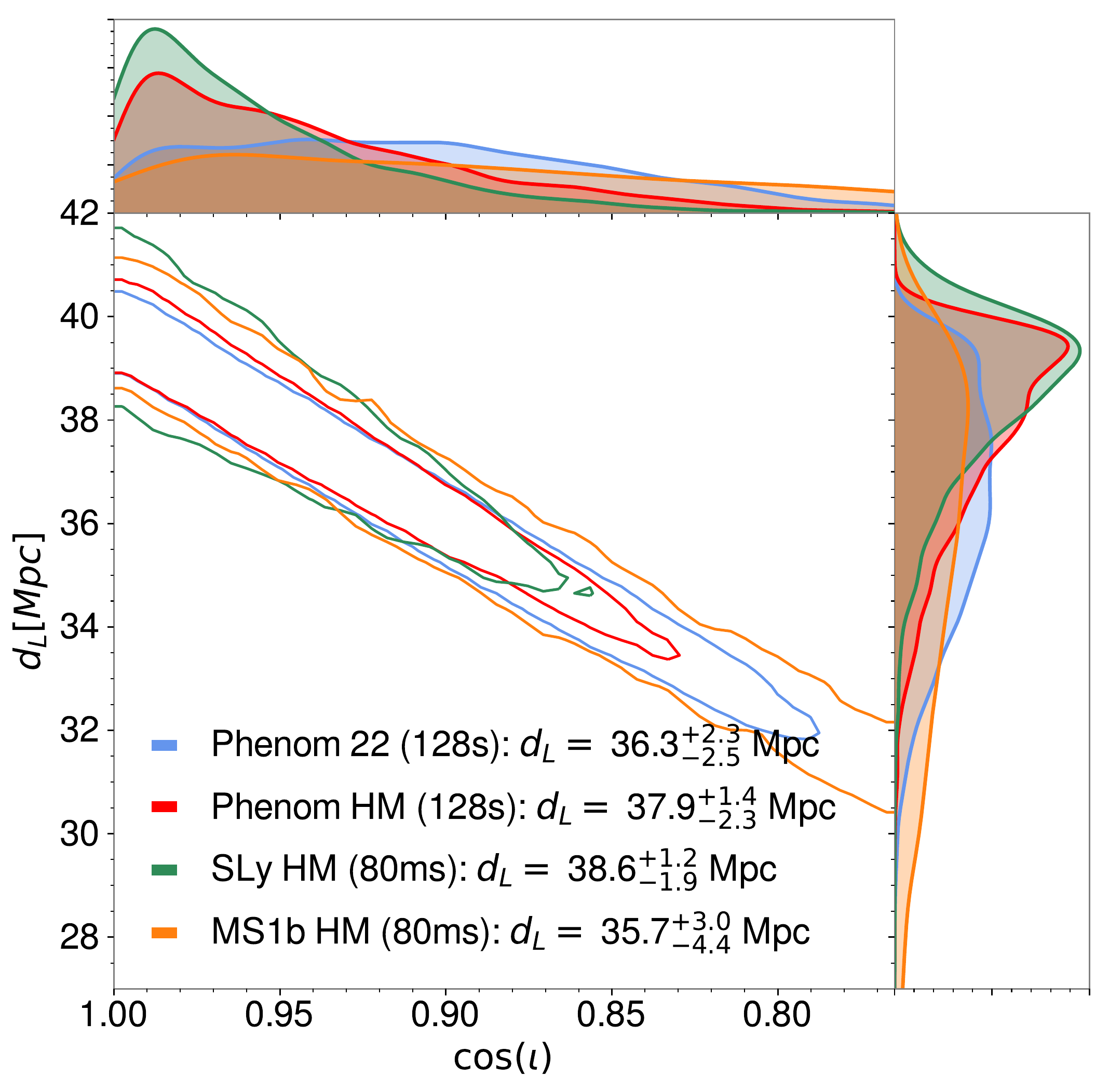}
\includegraphics[width=0.49\textwidth]{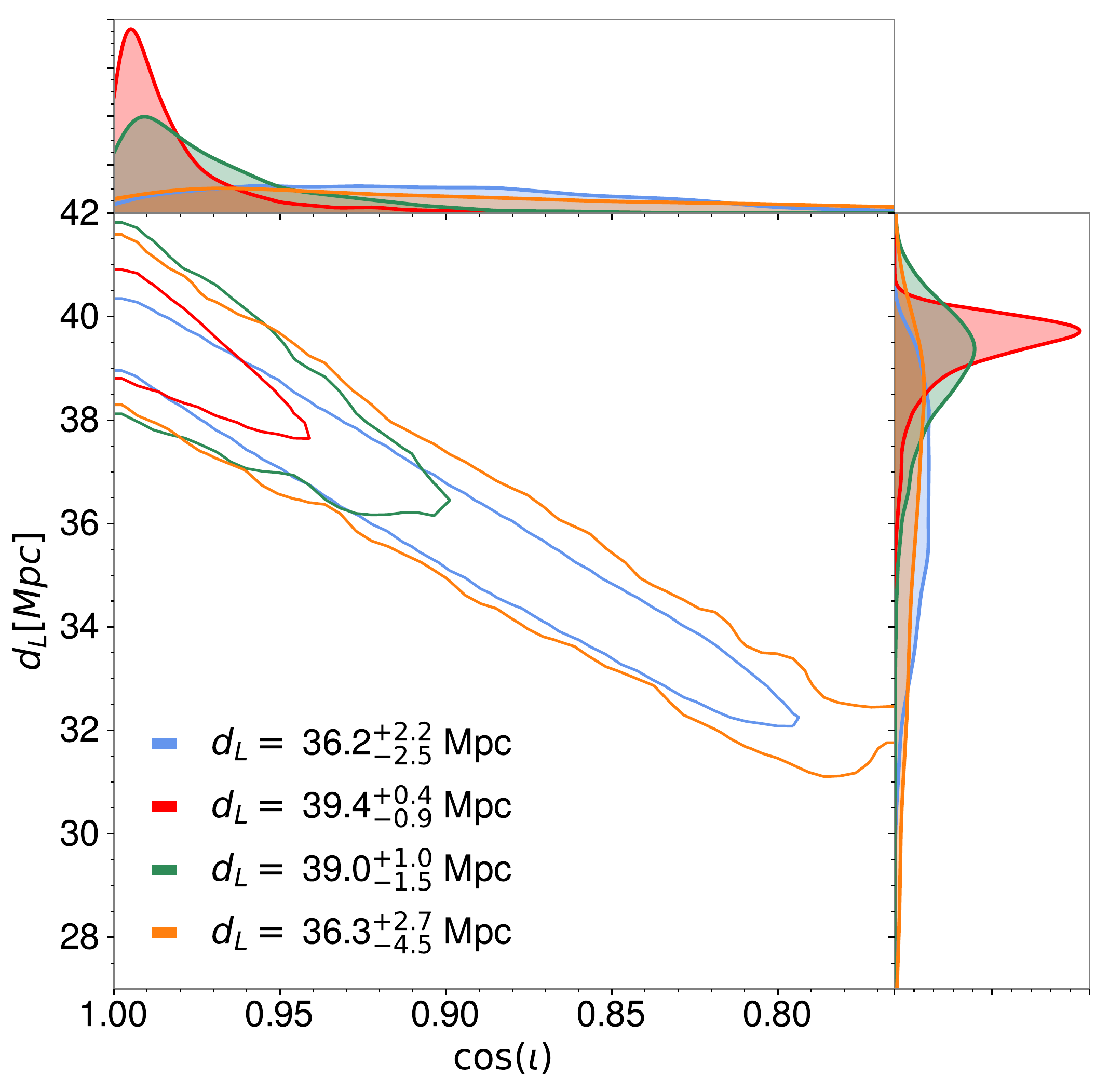}
\caption{\textcolor{black}{Two-dimensional posterior distributions for the luminosity distance and inclination angle of a face-on $(\cos(\iota)=1)$ binary neutron star merger located at 40 Mpc, with mass ratio $q=1$ (left) and $q=1.5$ (right) and a total mass of $M=2.75M_\odot$. The contours delimit the $90\%$ credible regions obtained by analysing a 128-s waveform including (red) and omitting (blue) higher-order modes. Also shown are posterior reconstructions using waveforms that only cover the last 80 ms of the merger and ringdown modelled with numerical relativity simulations with two different EOSs (orange and green). The labels quote median values and symmetric $68\%$ credible intervals for the luminosity distance.}} 
\label{fig:ID_ind}
\end{figure*}

\subsection{\textbf{Analysis setup}}

We inject signals $h(\theta_{true})$ with source parameters $\theta_{true}$, that include HMs, in zero noise and estimate the source parameters using waveform templates $h(\theta)$ that omit and include HMs. \textcolor{black}{We compute the  posterior Bayesian probability of the parameters $\theta$ as
\begin{equation}
 p(\theta|\theta_{true})=\frac{\pi(\theta){\cal{L}}(\theta|h(\theta_{true}))}{\int\pi(\theta){\cal{L}}(\theta|h(\theta_{true}))d\theta},   
\end{equation}
with $\pi(\theta)$ denoting the prior probability of the parameters $\theta$ and ${\cal{L}}(\theta|h(\theta_{true}))$ denoting their likelihood.} As usual, the latter is defined as the standard frequency-domain likelihood for GW transients \citep{Finn1992,Romano2017}
\begin{equation}
    \log{\cal{L}}(\theta|h(\theta_{true})) \propto -\sum_N\frac{(h(\theta_{true})-h(\theta)|h(\theta_{true})-h(\theta))}{2},
\end{equation}
%Here, $d$ denotes the detector data while $h(\theta)$ denotes a waveform template with parameters $\theta$. In our case $d$ is a simulated signal $h(\theta_{true})$ with source parameters $\theta_{true}$. %, which encompasses the intrinsic parameters $\Xi$, orientation $(\iota,\varphi)$, sky-location, signal polarisation and time of arrival. 
\textcolor{black}{where N runs over the different detectors of our network. As we discuss later, we work with two and three-detector networks.} As usual, $(a|b)$ represents the inner product \citep{Cutler1994}
\begin{equation}
   (a|b)= 4 \Re \int_{f_{min}}^{f_{max}} \frac{\tilde{a}(f)\tilde{b}^{*}(f)}{S_n(f)}df,
\end{equation}
where $\tilde{a}(f)$ denotes the Fourier transform of $a(t)$ and ${}^*$ the complex conjugate. The factor $S_{n}(f)$ denotes the one-sided power spectral density of the detector. In this work, we consider a network of detectors, all with noise sensitivity equivalent to that of the proposed 2.5-generation Neutron star Extreme Matter Observatory NEMO~\citep{OzHF}. We choose a lower frequency cutoff of $f_{min}=20$Hz and a sampling frequency of $16$kHz so that $f_{max}=8$kHz. The NEMO detector has a proposed sensitivity similar to Cosmic Explorer and Einstein Telescope in the kHz regime; we could equally use those third-generation detectors and achieve similar results for the late inspiral and post-merger, albeit with larger signal-to-noise ratios for the full signal given the better low-frequency ($\lesssim500$ Hz) sensitivity.

In all cases, we assume standard prior probabilities for the  sky-location, source orientation and polarisation angle, together with a prior uniform in co-moving volume and an uniform prior on the time-of-arrival, with a width of $0.2\rm \ s$, centered on the true value. For our analyses making use of 128s-long Phenom waveforms, we impose uniform priors on the individual masses $m_{1,2}\in[1,2]M_\odot$ and on the components of the individual spins along the orbital angular momentum $\chi^{z}_{1,2}\in[-0.15,0.15]$. Since numerical-relativity waveforms, however, are only produced for a discrete set of intrinsic parameters, we assume the masses and spins to be known in this case. We find this is a reasonable assumption as the individual masses and the effective spin parameter \citep{Santamaria:2010yb} $\chi_{eff}=(\chi^{z}_1 m_1 + \chi^{z}_2 m_2)/(m_1+m_2)$ are very well measured from the long inspiral. As an example, with a triple-detector network and using HMs, we determine the total mass, chirp mass and effective-spin parameters of our face-on unequal-mass source with respective uncertainties of  $<1\%$, $<0.01\%$ and $<0.015$ at the $68\%$ level. We perform our parameter inference runs with the software \texttt{Bilby}~\citep{Ashton:2018jfp,Parallel_Bilby}, sampling the parameter space with the algorithm \texttt{CPNest}~\citep{CPNest}.

We consider three network configurations.
The first (denoted HV) assumes two detectors with the location and orientation of Advanced LIGO Hanford and Virgo. Such a network has each detector sensitive to one of the two independent GW polarisations.
The second network (HL) assumes LIGO Hanford and Livingston location and orientations, almost anti-aligned, so that both detectors are sensitive to roughly the same GW polarisation, missing the other one. Finally, we consider an HLV network that is sensitive to both GW polarizations and can pinpoint the sky-location of the source.\\

\textcolor{black}{The accuracy of our distance measurement is of course limited by the loudness of the injected signals in our detector network, quantified by the optimal network signal-to-noise ratio (SNR), given by
\begin{equation}
\rho_{opt}(h(\theta_{true})) = \sqrt{\sum_N (h(\theta_{true})|(h(\theta_{true}))_{N}},
\end{equation}
which is inversely proportional to the source luminosity distance $d_L$. As we show in Appendix III, for face-on cases we obtain optimal SNRs of $\sim 190$ for HL and HLV networks and $\sim 140$ for HV. In contrast, for the weaker edge-on cases we obtain respective values $\rho_{opt} \in [30,65]$ depending on the mass ratio and network considered. For comparison, the SNR of GW170817, whose source was rather face-on, was  only $\simeq 32$. We note that while in this study we restrict to sources placed at distances $d_{L,true}=40$Mpc, our results can be extended to different reference distances. In particular, in our SNR-regime, given fixed source and detector network, the uncertainty of our distance estimates roughly depends on the optimal SNR (and therefore on the reference distance) as $ \Delta(d_L) \propto 1/\rho_{opt}^2 \propto d_{L,true}^2$ \citep{Cramer,Rao,Tjonniethesis}. In addition, uncertainties in the estimated value of the Hubble constant $H_0=cz/d_{L,true}$ grow linearly with distance.}

\subsection{Target binary neutron-star sources}
We choose four target sources with total total mass $M=2.75 M_\odot$, mass ratios $q=1$ and $q=1.5$, and oriented both face-on ($\iota=0$) and edge-on ($\iota=\pi/2$). Signals emitted in these two angles differ in three aspects: morphology, polarisation, and loudness. Face-on signals contain solely the $(\ell,m)=(2,2)$ mode, are circularly polarised (i.e., both $h_{+,\times}$ have the same amplitude), and are louder than edge-on signals. On the contrary, edge-on signals have contributions from higher-order emission modes that confer a richer structure, but are weaker in amplitude than face-on ones (see Fig.~\ref{fig:modes}). In addition, edge-on signals contain only one of the two polarisations. Consequently, it has been shown that measuring the ratio of the two polarisations is key to infer correctly the inclination of the source, provided that the detector network can observe both polarisations~\citep{Usman:2018imj}. For each of these sources, we consider two EOSs, namely \texttt{SLy} and \texttt{MS1b}, which we assume to be known. Different EOSs trigger post-merger HMs in different ways, varying the accuracy of the distance estimate. \textcolor{black}{Finally, we would ideally consider a wide range of distances and sky locations for all of our target sources. However, given the extreme computational cost of our parameter inference runs, and to allow for a direct comparison of our results, we place all of our sources at the same distance and sky location. For the former, we considered that the most reasonable choice was a value of $d_L=40$Mpc, consistent with that of GW170817, the only conclusive BNS observed to date through GWs. Finally, we placed all of our target sources at the same, arbitrary sky location.}\\

\begin{figure}
\includegraphics[width=0.48\textwidth]{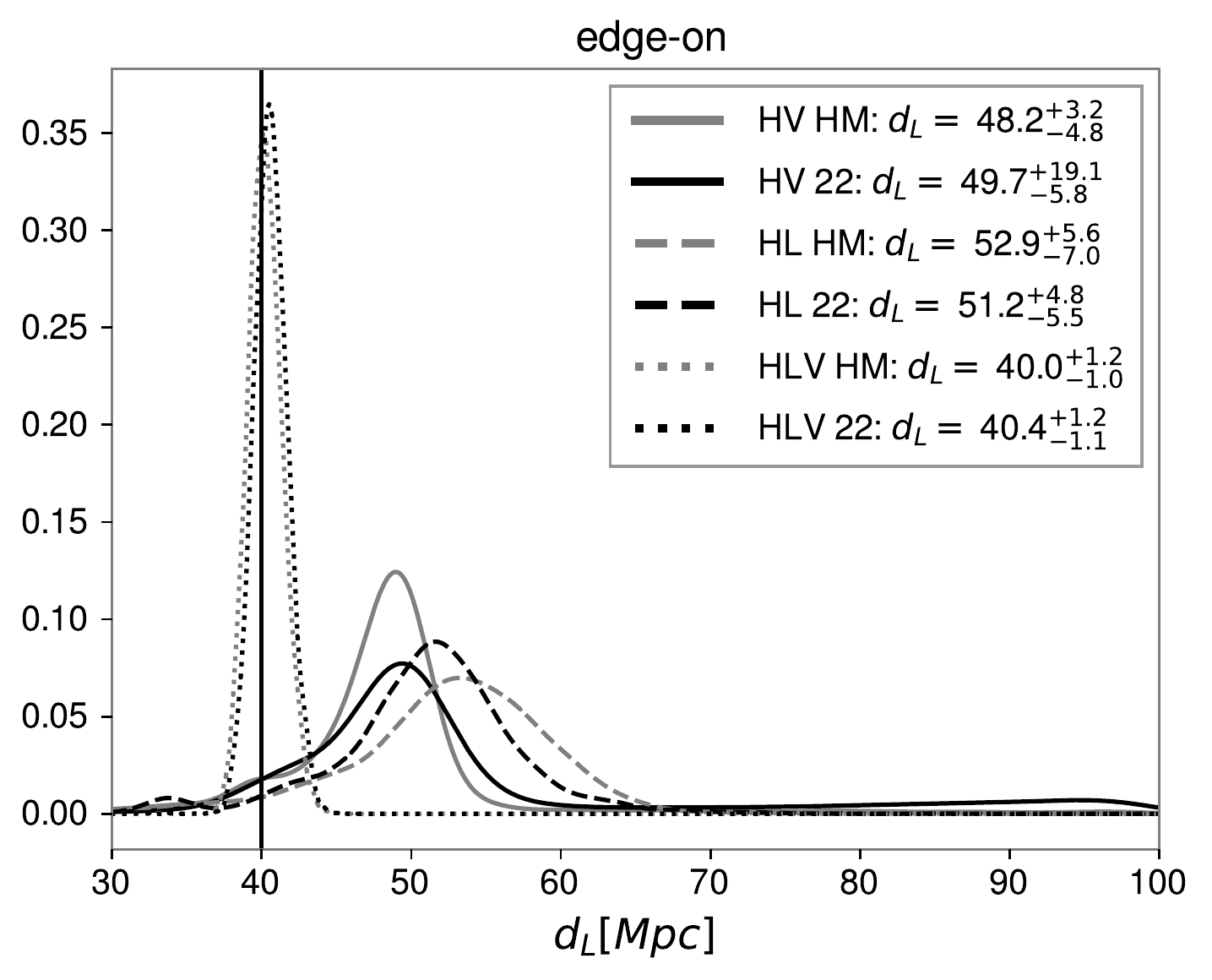}
\caption{Luminosity distance estimates for a binary neutron star merger with mass ratio $q=1.5$ and total mass $M=2.75M_\odot$ located at 40 Mpc using different detector configurations and omitting/including higher-order modes in the analysis of the signal. We quote median values and symmetric $68\%$ credible intervals. The injected signals are 128-s long IMRPhenomHM waveforms including higher-order modes.} 
\label{fig:edgeon}
\end{figure}

\begin{figure*}
\includegraphics[width=0.48\textwidth]{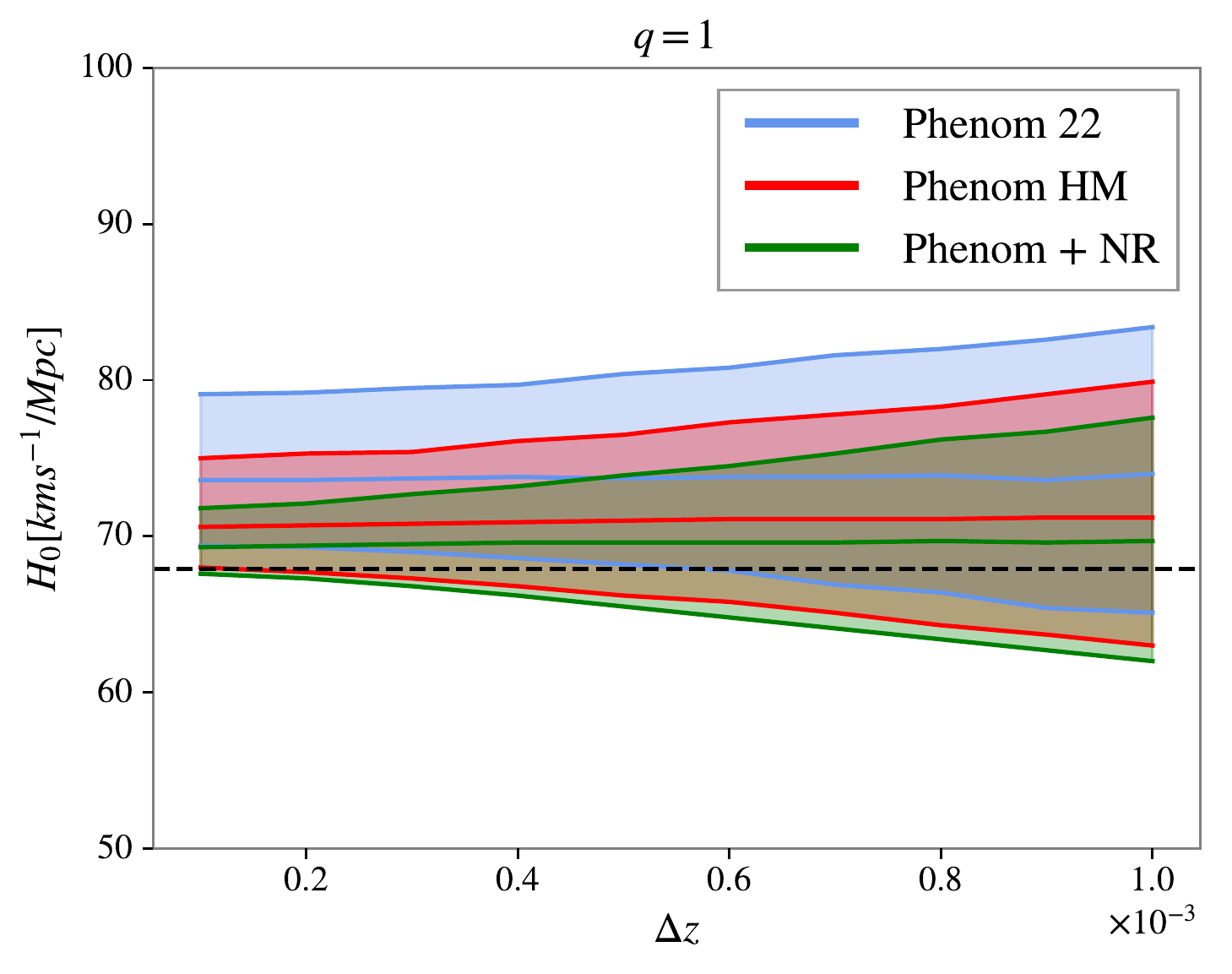}
\includegraphics[width=0.48\textwidth]{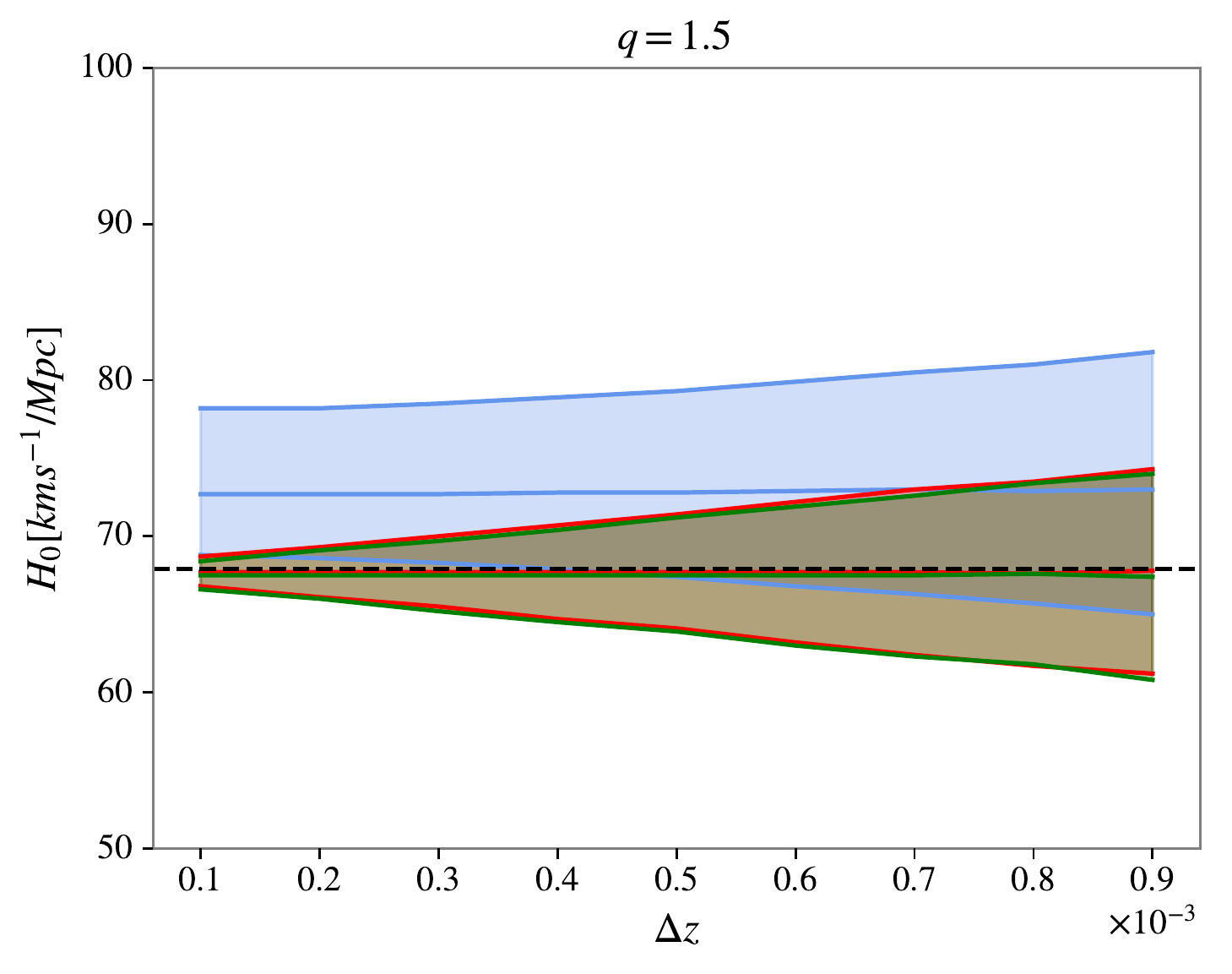}
\caption{Estimates of $H_0$ derived from the distance estimates of face-on binaries with mass ratios $q=1$ (left) and $q=1.5$ (right) located at a distance of 40 Mpc, and assuming redshift measurement of $z=0.00897 \pm \Delta z$. We consider a three-detector network in an HLV configuration with NEMO sensitivity curves. The contours delimit $68\%$ credible intervals. We show results using Phenom waveforms for binary black holes omitting and including HMs in blue and red, respectively. In green, we show combined results for Phenom and post-merger numerical-relativity waveforms including HMs.}
\label{fig:H0}
\end{figure*}

\section{Results}

\subsection{\textbf{Distance estimates}}

Figure~\ref{fig:ID_ind} shows the two-dimensional posterior distributions for the luminosity distance and inclination of face-on oriented binaries with mass ratios of $q=1$ (left) and $q=1.5$ (right), using an HLV detector configuration. The contours denote the $90\%$ credible regions and the legend provides median estimates with symmetric $68\%$ credible intervals. For unequal masses, and using 128s-long Phenom waveforms, the omission of HMs in the templates $h(\theta)$ leads to a biased estimate of $d_L^{\text{Phenom, 22}}=36.2^{+2.2}_{-2.5}$~Mpc. Their inclusion corrects this bias and reduces the uncertainty by $\sim 70\%$, yielding a distance measurement of $d_L^{\text{Phenom, HM}}=39.4^{+0.4}_{-0.9}$.
For equal masses, the impact of HMs is significantly milder, as the (usually strongest) odd-$m$ emission modes are suppressed \citep{Pan:2011gk,Pan:2013rra,Blanchet:2014zz}. This leads to a biased estimate of $d_L^{\text{Phenom, HM}}=37.9^{+1.4}_{-2.3}$ even if HMs are included. We also note that this remains true even if one would assume the intrinsic source parameters (masses and spins) to be known.
%\textcolor{blue}{Maybe highlight that this is the case even if the source parameters are fixed to the true ones}
This situation changes, however, when using $80ms$-long numerical-relativity waveforms that can account for the rich post-merger signal morphology. %The green and orange contours show estimates obtained with \texttt{SLy} and \texttt{MS1b} EOSs. 
For both mass ratios, and considering a \texttt{SLy} equation of state, results are better than those making use of 128s-long Phenom waveforms omitting HMs. Moreover, for equal-mass, the estimate even improves on that including HMs, yielding a non-biased estimate $d_L^{\texttt{SLy}}=38.6^{+1.8}_{-2.7}$ with smaller uncertainties.

We combine the distance estimates obtained using numerical-relativity waveforms with those obtained analysing Phenom waveforms restricted to frequencies not covered the former. To this, we multiply the respective posterior distributions for the the distance, dividing by one instance of the prior \citep{Zimmerman:2019wzo} \footnote{Note that this ignores that stronger constraints can be obtained for the inclination and the sky-location by combining both measurements, making our results rather conservative.}. We obtain joint estimates of $d_L^{\text{joint}} = 38.6^{+0.9}_{-1.3}$ Mpc for the $q=1$ case and $d_L^{\text{joint}} = 39.4^{+0.4}_{-0.7}$ for the $q=1.5$ case. The reason behind this improvement is that matter effects arising during the post-merger of BNSs trigger HMs, helping to break the degeneracy between distance and inclination and even allowing to measure the azimuthal angle (see Appendix II). For the unequal-mass case, these only add a small contribution with respect to the integrated effect of the HMs during the 128s of the signal. For $q=1$, however, odd-m modes are suppressed during the inspiral and are only triggered during the post-merger of our numerical-relativity simulations due to an effect known as \textit{one-armed spiral instability} or \textit{21-mode instability} \citep{East:2016zvv,Radice:2016gym,Lehner:2016wjg} (see Suppl.\ Material).
As a consequence, the inclusion of the post-merger emission can have a large impact. In our study, the distance estimates are much better for \texttt{SLy} compared to \texttt{MS1b}, due to the stronger HM emission. 

Finally, while we have discussed results assuming an HLV detector network, in the Suppl. Material we show results for all studied network configurations. In all cases we obtain similar results, both qualitatively and quantitatively.\\

\subsection{Edge-on cases}

Previous work has shown that the degeneracy between distance and inclination can be broken by computing the ratio $h_\times / h_+$, as this evolves from $1$ to $0$ as the inclination varies from face-on to edge-on \citep{Usman:2018imj}. We find that for face-on binaries, all HL, HV, and HLV configurations yield almost equivalent distance measurements despite the differences in signal loudness across the network, so that HMs have a much larger impact on the measurement than the polarisation ratio. In contrast, in Fig.~\ref{fig:edgeon} we show that for edge-on cases it is key not only to access both signal polarisations but also having a third detector that can pin-point the sky-location of the source. For an HLV configuration, unbiased estimates with uncertainties lower than $4\%$ are obtained regardless of the usage of HMs, while biased estimates are obtained using both two-detector configurations. The reason is that such configurations cannot pin-point the sky location of the source using timing information. This way, Bayesian inference places the source at those patches of the sky, consistent with the two-detector timing, where the detector network is most sensitive to, biasing the distance toward large values. We obtain identical qualitative results for the $q=1$ case, and when analysing our 80 ms-long numerical-relativity simulations.\\

\subsection{\textbf{Hubble constant estimates}}

Combining our distance estimates with simulated redshift estimates, we can infer $H_0$ via $H_0=c z / d_L$, with $c$ the speed of light. We assume redshift estimates consisting of Gaussian posterior distributions centered at a value of $z_0=0.00897$, corresponding to $d_L=40$~Mpc in a $\Lambda$CDM cosmology with Hubble parameter $H_0=67.9$~km~s$^{-1}$~Mpc$^{-1}$. We assume that the redshift has an uncertainty with standard deviation $\Delta z$ ranging from a realistic value of $10^{-3}$ consistent with that for GW170817 \citep{Abbott:2017xzu} to an improved value of $10^{-4}$ which would require higher resolution spectrograms and also the possibility to measure the internal motion of sources within individual galaxies, e.g.,~\citep{Davis:2019wet}.\\

Figure~\ref{fig:H0} shows the $H_0$ estimates derived from the distance measurements of our face-on BNSs with mass ratios $q=1$ and $q=1.5$ as a function of $\Delta z$ using an HLV network. Once again, we quote results in terms of symmetric $68\%$ credible intervals centred at the median value.
For unequal mass, and for $\Delta z = 10^{-3}$, the omission of HMs never leads to biased estimates. Their inclusion, however, leads an important improvement from $H_0=73.8^{+9.0}_{-9.8}$ to $H_0=68.3^{+7.5}_{-7.5}$. More spectacularly, for the most optimistic $\Delta z = 10^{-4}$ HMs improve the measurement from $H_0=73.8^{+5.3}_{-4.1}$ to $H_0=68.0^{+1.6}_{-1.0}$, enabling a $2\%$-level measurement. This shows that with 2.5G detectors, together with improved detector calibration and waveform models, $H_0$ estimates will be limited by EM redshift measurements and not by GW distance ones. Conversely, if HMs are omitted, a significant reduction of $\Delta z$ will \textit{not} translate into an improved $H_0$ estimate. Moreover, we find that $H_0$ estimates would be biased when $\Delta z \lesssim 0.5$. Consistently with the previous section, the inclusion of the post-merger emission in the analysis does not lead to any relevant improvement.

As expected, the situation is different for equal-mass cases. For these, the inclusion of post-merger effects is crucial to obtain visible improvements in the $H_0$ measurement. For $\Delta z = 10^{-3}$ we obtain a mild improvement from $H_0=71.2^{+8.6}_{-8.3}$ to $H_0=69.6^{+7.8}_{-7.7}$. When $\Delta z$ is reduced to $\Delta z = 10^{-4}$, the HMs present in the post-merger emission allow for an estimate $H_0=69.3^{+2.5}_{-1.5}$, with uncertainties at the $\simeq 4\%$ percent level, while their omission doubles the uncertainty and biases the measurement.\\

Almost identical results hold for the HL and HV networks, as shown in the Suppl.\ Material. For the (weaker) edge-on binaries, we find percent-level measurements are possible using the HLV network regardless of the usage of HMs.\\

\section{\textbf{Conclusions}}

We have shown that the use of higher-order modes in parameter inference of compact binaries with masses in the BNS range leads to great improvements of the distance and inclination estimates in the context of future detectors sensitive to signals in the $\sim$ kHz regime, such as the proposed 2.5-generation instruments presented in Refs.~\citep{Martynov2019,OzHF} and full third-generation interferometers~\citep{Punturo:2010zz,Reitze:2019iox}. For face-on binaries with modest mass ratios of $q=1.5$, we find that the accumulated effect of the higher-order modes during the inspiral reduces the uncertainties by $\approx70\%$. At the current state-of-the-art of redshift measurements from EM counterparts, this yields an $\approx25\%$ improvement of $H_0$ estimates. With improved redshift estimates, HMs can enable measurements of $H_0$ near the sub-percent level with a \textit{single observation}. For equal-mass binaries, the higher-order modes emitted during the post-merger stage are crucial to improve $H_0$ estimates. A soft EOS like \texttt{SLy}, favoured by current  observations, would enable percent-level measurements. For edge-on cases, we find that it is crucial to have a three-detector, HLV-like network, able to constrain the inclination and the sky-location of the binary.\\

We have focused on single event analyses, assuming a constant value for $H_0$. Significantly more precise measurements of the Hubble constant will be achievable by combining this method with an ensemble of binary neutron star detections in the not-too-distant future. The precision of the estimates we obtain with single events may enable us to study possible time variations~\citep{Wu1996} and anisotropies \citep{Collins1986} of $H_0$.\\ \textcolor{black}{While we have restricted to the reasonable paradigm of a GW170817-like source located at 40Mpc, we note that current BNS merger-rate estimates suggest that such an event is rather rare, with less than $ 0.1$ such mergers per year \citep{O2rates}. Pushing this value to at least 1 per year would require the consideration of sources at $\simeq 90$Mpc, halving the SNR of our signals, multiplying by four our distance uncertainties and doubling that for $H_0$. We note however that this would still allow for $4\%$-level measurements of $H_0$. Furthermore, and most importantly,} we have considered the fairly conservative scenario of a network formed by 2.5-generation detectors like NEMO, which resulted in signal-to-noise ratios (SNRs) ranging in $50-200$ (see Appendix III). The replacement of two of these detectors by other projected detectors like Cosmic Explorer \citep{CE,CE2} or Einstein Telescope \citep{ET1,ET2}, more sensitive at low frequencies and as sensitive at high frequencies as NEMO, does rise these SNRs to the order of 1000, which would lead to significantly more precise results.\\

\acknowledgments

We thank Rory Smith, Avi Vajpeyi and Sylvia Biscoveanu for their help to set up \texttt{Parallel Bilby}~\citep{Parallel_Bilby} runs. We thank Nicolas Sanchis-Gual, Ornella Piccinni, Valentin Christiaens and Ricardo Martinez-Garcia for comments on the manuscript and Tjonnie Li for useful discussions. JCB acknowledges support by the Australian Research Council (ARC) Discovery Project DP180103155 and the Direct Grant, Project 4053406, from the Research Committee of the Chinese University of Hong Kong. The project that gave rise to these results also received the support of a fellowship from ”la Caixa” Foundation (ID
100010434) and from the European Union’s Horizon
2020 research and innovation programme under the
Marie Skłodowska-Curie grant agreement No 847648.
The fellowship code is LCF/BQ/PI20/11760016. PDL is supported through ARC Future Fellowship FT160100112, ARC Discovery Project DP180103155, and ARC Centre of Excellence CE170100004. The authors acknowledge computational
resources provided by the LIGO Laboratory and supported by
National Science Foundation Grants PHY-0757058 and PHY0823459; and the support of the NSF CIT cluster
for the provision of computational resources for our parameter inference runs, and the support of the CUHK Central High Performance Computing Cluster, on which our runs using Parallel Bilby were performed. Part of the work described in this paper was supported by a grant from the Croucher Foundation of Hong Kong. This document has LIGO DCC number LIGO-P2000160.\\

\begin{figure}[h]
\includegraphics[width=0.49\textwidth]{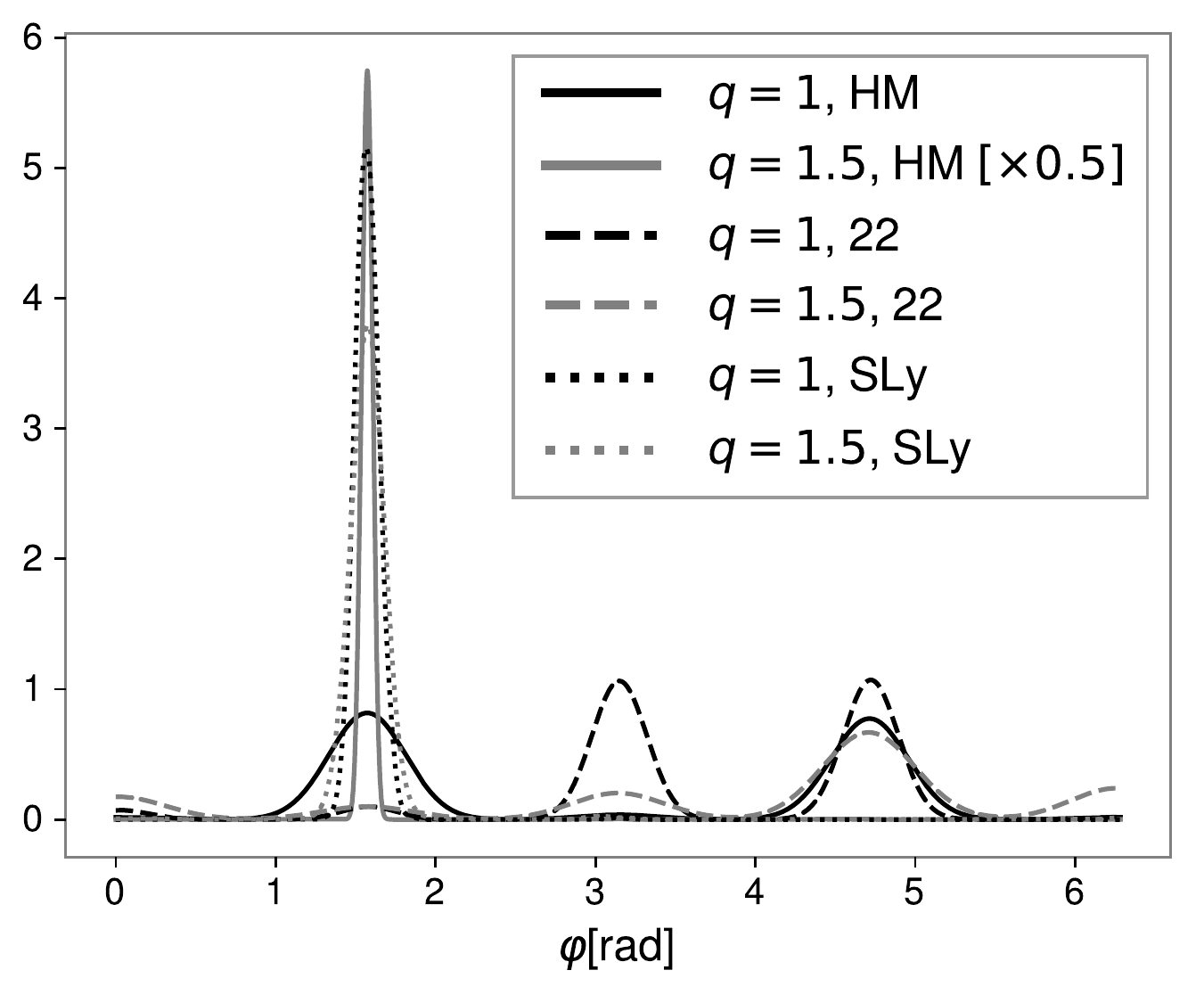}
\caption{\textbf{Estimation of the azimuthal angle} for two edge-on compact binaries with mass ratios $q=1$ (black) and $q=1.5$ (grey). The solid and dashed lines denote estimates making use of 128s-long Phenom models for BBHs, respectively including and omitting HMs. The dotted lines denote estimates based on 80ms-long BNS simulations including HMs. For $q=1$, only the usage of numerical-relativity waveforms allows for an unambiguous estimation of the azimuth.} 
\label{fig:phi}
\end{figure}

\section*{\textbf{Appendix I: Measuring the azimuthal angle}}
The inclusion of HMs in the templates accounts for asymmetries in the GW emission that cannot be reproduced when HMs are omitted. This asymmetry allows to define preferred directions in the orbital plane to determine in an unambiguous way where the observer is sitting on it, providing a clear physical meaning to the azimuth angle $\varphi$ \citep{CalderonBustillo:2019wwe}. For instance, for the case of BBHs, this was used to determine the direction of the final recoil velocity with respect to the line of sight \citep{CalderonBustillo:2018zuq}.  While this would lead to accurate measurements of the kick introduced by GW radiation \citep{CalderonBustillo:2018zuq}, it might not account for the final velocity of the remnant due to the 
ejection of material during the merger process, e.g.,~\citep{Kyutoku:2015gda,Chaurasia:2020ntk}. Fig.~\ref{fig:phi} shows the posterior distributions for the $\varphi$ for two edge-on binaries with mass ratios $q=1$ and $q=1.5$, using a ``HLV'' network. Results are shown for 128s-long Phenom waveforms including and omitting HMs, and for 80 ms-long numerical-relativity waveforms implementing an \texttt{SLy} EOS and including HMs. For unequal masses, HMs allow for an extremely accurate estimate without the need to include post-merger effects. For equal-mass systems, however, this kind of measurement is only possible once the post-merger is included. In concordance with the results shown in Figs.~\ref{fig:ID_ind} and \ref{fig:H0}, the measurement is less precise for \texttt{MS1b}, due to its weaker HMs.\\

\section*{\textbf{Appendix II: Exploiting the one-armed inspiral instability} }

For the case of BBHs, HMs have very little contribution to the GW emission when the mass ratio is close to one. Thus, the large effect of the HMs for the measurement of the Hubble constant might seem somewhat unexpected. Moreover, Fig.~\ref{fig:modes} shows that the contribution from the HMs in the postmerger phase is dominated by that of the $(2,1)$-mode, which is particularly surprising for the case of the equal-mass binary, for which odd-m modes are completely suppressed for black hole binaries by the symmetry of the problem. Past work has shown that tiny asymmetries in the binary configuration can develop into a large asymmetry in the merger stage known as \textit{one-armed spiral instability}, triggering a strong $(2,1)$ mode, e.g., \citep{East:2016zvv,Radice:2016gym,Lehner:2016wjg}. Fig.~\ref{fig:modes} shows that this leads to ``kinks'' in the GW spectrum at a frequency of approximately half of the main emission peak. This feature grows as the binary is observed edge-on and its exact morphology depends strongly on the azimuthal angle. The importance of the $(2,1)$ mode suggests that we are in fact exploiting the one-armed spiral instability to infer the orientation of the binary. To check this, we repeated our analyses using numerical-relativity waveforms including only the $(\ell,m)=(2,2)$ and $(2,1)$ modes obtaining results quantitatively identical to those including  the rest of modes.\\

\section*{\textbf{Appendix III: Tables of estimates for face-on binaries.} }

In Tables \ref{table:q1fo} and \ref{table:q15fo} we report all luminosity distances and $H_0$ estimates obtained for face-on cases using Phenom waveforms including and omitting HMs, and combining Phenom and numerical-relativity estimates assuming an \texttt{SLy} EOS. \textcolor{black}{In order to provide an idea of the loudness of these injections, we quote optimal SNRs for those performed using the PhenomHM model, adding those for the edge-on cases within a parenthesis. For comparison, note that the SNR of GW170817, rather face-on, was of only $\simeq 32$, much lower than that of our weak edge-on injections}. As throughout the main body of the paper, we quote median values and symmetric $68\%$ credible intervals. For our edge-on binaries, as shown in Fig.~\ref{fig:edgeon}, two detector networks produce extremely biased results and measurements are independent of the usage of HMs.

\begin{sidewaystable}[]
\begin{tabular}{c c c c  c c c  c c c}
%\toprule
\\[1pt]
Waveform Model &  \multicolumn{3}{c}{$d_L\,[\mathrm{Mpc}]$} & \multicolumn{3}{c}{$H_0\,[\mathrm{km\; s^{-1}/Mpc}]\ (\Delta z = 10^{-3})$} & \multicolumn{3}{c}{$H_0\,[\mathrm{km\; s^{-1}/Mpc}]\ (\Delta z = 10^{-4})$}\\
\hline
{}   & HLV   & HV    & HL  & HLV   & HV    & HL & HLV   & HV & HL \\   
\hline
\textcolor{black}{Optimal SNR}  & 198.67 (63.58) & 141.18 (52.97) & 189.17 (56.96) & & & & &\\ \hline \\ [3pt]
Phenom $(2,2)$  & $36.3^{+2.3}_{-2.5}$ & $35.7^{+2.4}_{-3.7}$ & $32.8^{+5.0}_{-6.5}$  & $73.9^{+9.6}_{-8.8}$ & $74.3^{+10.4}_{-9.3}$ & $81.8^{+18.9}_{-13.1}$ & $73.6^{+5.4}_{-4.2}$ & $72.8^{+7.3}_{-4.8}$ & $81.2^{+18.4}_{-10.2}$  \\
\\[6pt]
Phenom HM   &  $37.9^{+1.4}_{-2.3}$  & $38.1^{+1.4}_{-2.8}$  & $37.5^{+1.8}_{-3.1}$ &  $71.2^{+8.8}_{-8.3}$    & $73.2^{+9.5}_{-9.0}$ & $72.2^{+9.0}_{-7.8}$ & $70.6^{+4.4}_{-2.5}$ & $72.5^{+6.1}_{-4.0}$ & $71.4^{+5.9}_{-3.3}$ \\
\\[6pt]
Phenom HM + NR[\texttt{SLy}] $\; \;$  &  $38.6^{+0.9}_{-1.3}$ & $38.4^{+1.0}_{-1.7}$ & $38.4^{+1.0}_{-2.2}$ &  $69.6^{+7.8}_{-7.7}$ &  $69.8^{+7.1}_{-7.5}$ & $70.3^{+8.0}_{-7.8}$ & $69.3^{+2.5}_{-1.5}$ & $69.3^{+1.9}_{-2.9}$ & $69.8^{+3.2}_{-2.0}$ \\
\hline 
\end{tabular}
\caption{Luminosity distance and $H_0$ estimates for a face-on binary with mass ratio $q=1$ and total mass $M=2.75 M_\odot$ located at a true distance of $40$\ Mpc using different waveform models and detector network configurations. We quote median values and symmetric $68\%$ credible intervals. \textcolor{black}{On the first row, we provide the optimal network SNRs of the injections performed with PhenomHM. For completeness, we also add within a parenthesis the value for the edge-on versions, which is significantly lower.}}
\label{table:q1fo}
\end{sidewaystable}

\begin{sidewaystable}
\begin{tabular}{c c c c  c c c  c c c}
%\toprule
\\[1pt]
Waveform Model &  \multicolumn{3}{c}{$d_L\,[\mathrm{Mpc}]$} & \multicolumn{3}{c}{$H_0\,[\mathrm{km\; s^{-1}/Mpc}]\ (\Delta z = 10^{-3})$} & \multicolumn{3}{c}{$H_0\,[\mathrm{km\; s^{-1}/Mpc}]\ (\Delta z = 10^{-4})$}\\
\hline 
{}   & HLV   & HV    & HL  & HLV   & HV    & HL & HLV   & HV & HL \\   
\hline

\textcolor{black}{Optimal SNR} & 194.44 (62.23) & 149.14 (34.40) & 185.15 (55.75) & & & & & \\ \hline \\ [3pt]

Phenom $(2,2)$  & $36.2^{+2.2}_{-2.5}$ & $36.0^{+2.5}_{-3.4}$ & $34.2^{+3.7}_{-5.1}$  & $73.8^{+9.0}_{-9.8}$ & $74.8^{+12.3}_{-9.3}$ & $77.0^{+11.7}_{-10.6}$ & $73.8^{+5.3}_{-4.1}$ & $74.2^{+7.5}_{-4.7}$ & $78.1^{+13.3}_{-7.5}$ \\
\\[6pt]
Phenom HM   &  $39.4^{+0.4}_{-0.9}$  & $38.9^{+0.8}_{-2.1}$  & $38.5^{+1.1}_{-2.3}$  &  $68.3^{+7.5}_{-7.5}$    & $69.5^{+8.4}_{-7.8}$ & $74.1^{+10.1}_{-9.2}$ & $68.0^{+1.6}_{-1.1}$ & $68.9^{+3.6}_{-1.7}$ & $69.5^{+4.2}_{-2.1}$ \\
\\[6pt]
Phenom HM + NR[\texttt{SLy}] $\; \;$  &  $39.4^{+0.4}_{-0.7}$ & $38.9^{+0.8}_{-2.3}$ & $39.0^{+0.7}_{-1.1}$ &  $68.1^{+7.2}_{-7.3}$ & $69.0^{+7.8}_{-7.7}$ & $67.4^{+5.8}_{-5.7}$ & $68.0^{+1.2}_{-1.0}$ & $68.7^{+1.9}_{-1.4}$ & $68.8^{+2.3}_{-1.5}$ \\ 
\hline 
\end{tabular}
\caption{Luminosity distance and $H_0$ estimates for a face-on binary with mass ratio $q=1.5$ and total mass $M=2.75 M_\odot$ located at a true distance of 40~Mpc using different waveform models and detector network configurations. We quote median values and symmetric $68\%$ credible intervals. \textcolor{black}{On the first row, we provide the optimal network SNRs of the injections performed with PhenomHM. For completeness, we also add within a parenthesis the value for the edge-on versions, which is significantly lower.}}
\label{table:q15fo}
\end{sidewaystable}

\bibliography{IMBBH.bib}{}
\bibliographystyle{aasjournal}

\end{document}